\documentclass[12pt,preprint]{aastex}

\def\be{\begin{equation}}
\def\ee{\end{equation}}
\def\ba{\begin{eqnarray}}
\def\ea{\end{eqnarray}}
\newcommand{\degree}{\ensuremath{^\circ}}

\newcommand{\sgr}{Sgr\,A*}
\newcommand{\msun}{\ifmmode\mbox{M}_{\odot}\else$\mbox{M}_{\odot}$\fi}
\newcommand{\rsun}{\ifmmode\mbox{R}_{\odot}\else$\mbox{M}_{\odot}$\fi}
\newcommand{\degrees}{\ifmmode^{\circ}\else$^{\circ}$\fi}
\newcommand{\amin}{\ifmmode^{\prime}\else$^{\prime}$\fi}
\newcommand{\asec}{\ifmmode^{\prime\prime}\else$^{\prime\prime}$\fi}

\shorttitle{Three Pulsars Near \sgr}
\shortauthors{Deneva et al.}

\begin{document}

\title{Discovery of Three Pulsars from a Galactic Center Pulsar Population} 
\author{
J.~S.~Deneva$^{1,*}$, 
J.~M.~Cordes$^{1}$,
and T.~J.~W.Lazio$^{2}$}

\affil{$^{1}$Astronomy Dept., Cornell Univ., Ithaca, NY 14853}
\affil{$^{2}$Naval Research Laboratory, 4555 Overlook Avenue, SW, Washington, DC 20375}

\begin{abstract}
We report the discovery of three pulsars whose large dispersion measures and angular proximity to \sgr\ indicate the existence of a Galactic center population of neutron stars. The relatively long periods (0.98 to 1.48~s) most likely reflect strong selection against short-period pulsars from radio-wave scattering  at the observation frequency of 2~GHz used in our survey with the Green Bank Telescope. One object (PSR J1746-2850I) has a characteristic spindown age of only 13~kyr along with a high surface magnetic field $\sim 4\times 10^{13}$~G. It and a second object found in the same telescope pointing, PSR J1746-2850II (which has the highest known dispersion measure among pulsars), may have originated from recent star formation in the Arches or Quintuplet clusters given their angular locations.  Along with a third object, PSR J1745-2910, and two similar high-dispersion, long-period pulsars reported by \cite{j+06}, the five objects found so far are 10 to 15~arc~min from \sgr, consistent with there being a large pulsar population in the Galactic center, most of  whose members are undetectable in relatively low-frequency surveys because of pulse broadening from the same scattering volume that angularly broadens \sgr\ and OH/IR masers.   
\end{abstract}     


\section{Introduction}

The scientific pay-offs from finding pulsars orbiting near \sgr\ 
are potentially very high and fall into three main categories. Measurements 
of relativistic effects through timing of pulsars in tight 
orbits around \sgr\ would provide methods for better constraining the 
mass of the central black hole and even estimating its spin (e.g. \citealt{lw97}, \citealt{wk99}, \citealt{pl04}). Timing measurements will also characterize the distribution of dark matter near Sgr A* either in the form of a cluster of black holes and neutron stars or in a smoothly distributed volume containing dark-matter particles (\citealt{bm05}, \citealt{Weinberg05}).
The spatial, age and period distributions of pulsars near \sgr\ will
help describe the stellar population in the region and discriminate between 
hypotheses attempting to explain the presence of the central 
cluster of young massive stars: stellar collisions and mergers, 
migration, and a past episode of intensive star formation. 
Finally, using pulsars to probe the scattering region around \sgr\ will lead to refinement of electron density models for the inner Galaxy.  

There are at least three stellar clusters in the Galactic 
center region containing massive stars, the progenitors of radio 
pulsars. \cite{g+03} present evidence that the central cluster 
around \sgr\ includes early-type stars.
Since stars with masses of about
$8-20\ \msun$ are neutron star 
(NS) progenitors, it is plausible that a considerable NS population exists 
as well. 
This leads to the conclusion that a sizable fraction 
of the NSs would be active radio pulsars. \cite{pl04} estimate 
that there are $\sim100-1000$ active pulsars orbiting \sgr\ with periods of less than 100 years. 
\cite{cl97} show that some of these pulsars will be detectable at 
the distance of the Galactic center, even at the high frequencies needed to 
mitigate pulse broadening from scattering. 

Apart from the central star cluster around \sgr, the Arches and Quintuplet clusters are the densest stellar associations within 0.5\degrees\ of the Galactic center. Both clusters are composed mainly of young massive stars whose supernova explosions will have produced neutron stars young enough to be active radio pulsars. \cite{Stolte08} show that the radial distance from the Galactic center to the Arches cluster is $62 \pm 23$~pc if the cluster is in a circular orbit around \sgr, but 
its distance may be as large as 200~pc.

Of the nearly 2000 pulsars known up to now, only two are within 
1\degrees\ of \sgr: PSR~J1746$-$2856 and PSR~J1745$-$2912, which were 
discovered by \cite{j+06} in a 3.1~GHz survey with the 
Parkes telescope and have dispersion measures (DMs) 
of~1168 and~1130~pc~cm${}^{-3}$, respectively. 
\cite{j+06} also report the absence of detections by a 
8.4~GHz Parkes survey, which indicates there are not many 
pulsars with flat enough spectra to be detected at high frequencies.
The non-detections are also  consistent with electron-density models that 
predict that scattering at frequencies up to $\sim 10$~GHz 
strongly affects the detectability of pulsars near \sgr\
\citep{cl97}. 

As part of a larger program to probe the Galactic center pulsar population and ionized gas environment, we have surveyed the inner 0.5\degrees\ around \sgr\ at 2~GHz with the Green Bank telescope. Pulse broadening due to scattering varies with frequency approximately as $\tau_s \propto \nu^{-4}$, so the broadening expected at 2~GHz is $\sim 5$ times larger than at $3.1$~GHz. This effect is partially mitigated by the fact that most pulsars have moderate to steep spectra and are more easily detected at lower frequencies when scattering is
not a factor.

\begin{figure}
\includegraphics[angle=0,width=\linewidth]{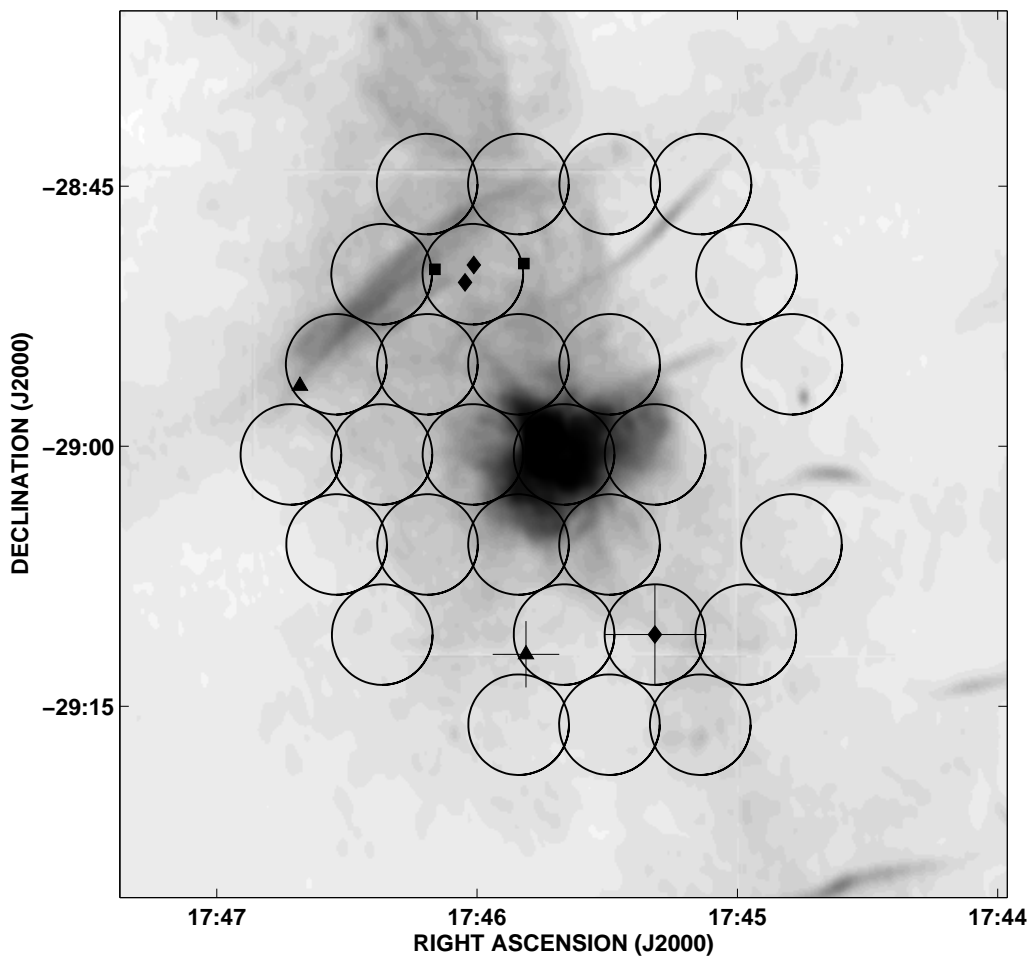}
\caption{
The grid of 2~GHz survey pointings from 2007 overlaid on a 0.33~GHz image of the Galactic center \citep{lklh00}. Circles correspond to the 5.8\amin\ FWHM beam size of the Green Bank telescope at 2~GHz. Diamonds denote the positions of J1746$-$2850I, J1746$-$2850II, and J1745$-$2910, and triangles denote the positions of J1746$-$2856 and J1745$-$2912. Crosses show actual position uncertainties. For J1746$-$2850I, J1746$-$2850II, and J1746$-$2856 the position uncertainties are smaller than the marker size. Squares show the positions of the Arches and Quintuplet clusters.\label{fig_grid}}
\end{figure}

\section{Survey Parameters and Data Processing}

The survey was carried out in September 2007 with the Green Bank telescope, using a center frequency of 1.95~GHz and the SPIGOT backend with a bandwidth of 800~MHz. The lower 200~MHz of the receiver band was excluded from the analysis due to the presence of severe radio frequency interference. The effective analyzed bandwidth was 600~MHz divided into 768 channels. Data were taken with a sampling time of 82~$\mu s$ in two polarizations, which were summed before further processing. The grid of survey pointings around \sgr\ is shown in Fig.~\ref{fig_grid} overlaid on a 330~MHz image of the Galactic center region. We observed 37 grid pointings for one hour each, but data from 8 pointings were not usable due to technical issues and these pointings are omitted from the figure. 

Data processing was performed at the Cornell Center for Advanced Computing (CAC) using the Cornell pulsar search code\footnote{\texttt{http://arecibo.tc.cornell.edu/PALFA/}} and the Presto package\footnote{\texttt{http://www.cv.nrao.edu/$\sim$sransom/presto}}. 
The full-resolution data were searched with 1245 evenly spaced trial DM values in the range 10 - 2000~pc~cm$^{-3}$, and 155 evenly spaced values in the range 2000 - 2500~pc~cm$^{-3}$. Data were also searched over 1000 values from 100 to 2100~pc~cm$^{-3}$ with time resolution of 0.65~ms and 300 values from 2100 to 3000~pc~cm$^{-3}$ with 1.3~ms resolution. A Fast Fourier Transform was performed on the resulting time series and the power spectrum was searched for periodic signals by summing up to 16 harmonics and applying a harmonic sum threshold of $6\sigma$. 

Time series were also searched for individual dispersed pulses using
two algorithms, one that applies simple templates (Cordes \& McLaughlin 2003)
 and another that searches for clusters of related points
(Cordes et al. 2004) using a friends-of-friends group-finding algorithm (see Fig.1 in \citealt{hg82}; see \citealt{Deneva08} for detailed description of application to single pulse searching). 
The matched filtering approach smoothes each time series with rectangle
functions of widths $2^n$ samples, where $n = 0 - 7$, and searches
for events above a 5$\sigma$ threshold. It is most sensitive to pulses 
whose widths are close to one of the boxcar filter widths, for a maximum 
width of 10~ms. The cluster algorithm 
identifies individual samples above a 3$\sigma$ threshold and then
combines contiguous points into events to which the larger 5$\sigma$ threshold
is applied.  


\section{Survey Results}

Table~\ref{tab_pulsars} summarizes the properties of the three discovered
pulsars. 
PSR J1746$-$2850I and PSR J1746$-$2850II were discovered in the same 
survey pointing and were confirmed with the Green Bank Telescope in June 2008; 
subsequent monthly timing observations at 2~GHz are ongoing. 
We observed both pulsars at 1.5 and 4.8~GHz with 
the dual goals of estimating their spectral indices and improving their 
position measurements. We also observed PSR~J1946$-$2850I at 9~GHz in order to confirm that its relatively flat spectrum extends to higher frequencies. 
PSR J1745$-$2910 was discovered in February 2009 and will be timed starting
in June 2009. Profiles of the three pulsars are shown in Figure~\ref{fig_profs}.

\begin{table}
\caption{Three New Pulsars Toward the Galactic Center\label{tab_pulsars}}
\begin{center}
\begin{tabular}{lccc}
\tableline
Parameter & J1746-2850I & J1746-2850II & J1745-2910 \\
\tableline
RA (J2000) & 17$^h$~46$^m$~06.6$^s$(2) &  17$^h$~46$^m$~03.7$^s$(1) & 17$^h$~45$^m$~16$^s$(34) \\
DEC (J2000) & -28\degrees~50\amin~42\asec(5) &  -28\degrees~49\amin~19\asec(21) & -29\degrees~10'(3)\\
$P$ (s) & 1.0771014910(4) & 1.478480373(2) &  0.982 \\
$\dot{P}$~(s/s) & 1.34311(2) $\times 10^{-12}$ & 1.27(6) $\times 10^{-14}$ & \\
$DM$ (pc~cm${}^{-3}$) & 962.7(7) & 1456(3) & 1088\\
Age (Myr) & 0.013 & 2 & \\
B (G) &  $3.8\times 10^{13}$ & $2.8\times 10^{12}$ & \\
$\dot{E}$ (erg/s)$^a$ & $4.24 \times 10^{34}$ & $1.47 \times 10^{32}$ & \\
\\
$W_{8.9 }$ (50\%,ms)$^b$ &  50  & & \\
$W_{4.8 }$ (50\%,ms) &  30  & 31 & \\
$W_{1.95}$ (50\%,ms) & 45  & 130 & 54 \\
$W_{1.5 }$ (50\%,ms) &  100  & 145 & \\
\\
$S_{8.9 }$ (mJy) &  0.4  & & \\
$S_{4.8 }$ (mJy) &  0.5  & 0.1 & \\
$S_{1.95}$ (mJy) & 0.6  & 0.2 & 0.2 \\
$S_{1.5 }$ (mJy) &  0.8  & 0.4 & \\
Spectral index $\alpha$ & -0.3 & -1.1 & \\
$(S_{\nu} \propto \nu^{\alpha})$ & & & \\
\tableline
\end{tabular}
\end{center}
$^a$ Assuming a 1.4~\msun\ neutron star with a 10~km radius and moment of inertia $I = 10^{45}$~g~cm$^{-3}$.
$^b$ Subscripts refer to observing frequencies in GHz.
\end{table}

We estimated the period-averaged flux density for each pulsar
by scaling from signal-to-noise ratio of the folded pulse profile 
$\left(S/N \right)_{\rm prof}$ 
using
the calculated radiometer noise level for the sum of two polarization
channels:  
\begin{equation}
S = \frac{\left(S/N \right)_{\rm prof}T_{\rm sys}}{G \sqrt{N_{\rm pol}~\Delta\nu~T_{\rm obs}/n_{\rm bin}}}, \label{eqn_flux}
\end{equation}
and we have used $T_{\rm sys}$ of 32~K, 27~K, 19~K, and 27~K at
1.4, 2, 4.8, and 9~GHz, including sky background contributions in
the Galactic center direction of 12~K at 1.4~GHz scaled as
$\nu^{-2.5}$ \citep{RR}.   
The gains $G$ are 
2.0~K~Jy$^{-1}$,
1.9~K~Jy$^{-1}$,
2.0~K~Jy$^{-1}$,
1.8~K~Jy$^{-1}$,
respectively at the four frequencies. The number of bins in the pulse profile
$n_{\rm bins} = 128$. The total bandwidth $\Delta\nu = $ 600~MHz at 1.4 and 2~GHz, and 800~MHz at 4.8 and 9~GHz.



No isolated dispersed pulses were detected above a $5\sigma$ threshold in any of the 2~GHz survey pointings. For a 100~ms pulse width (comparable to the observed pulse widths from Table~\ref{tab_pulsars}), we derive an upper limit on the flux density of single pulses
\be
S_{\rm max} = \frac{m T_{\rm sys}}{G \sqrt{N_{\rm pol}~\Delta\nu~W}} \approx 6.5~{\rm mJy}\label{eqn_sp}
\ee

\subsection{PSR J1746$-$2850I}

A timing solution for J1746-2850I was obtained from observations with
the GBT made at 24 epochs between June 2008 and August 2009 and fitted
for period $P$, period derivative $\dot P$, DM, and sky coordinates. 
The long period and
large period derivative indicate that this is a young
object with a characteristic age $\tau_{c} = P/2\dot{P} = 13$~kyr and 
large surface magnetic 
field,  $3.8\times10^{13}$~G. 
Assuming a power-law spectrum 
$S_\nu \propto \nu^{\alpha}$, we found that J1746-2850I has a 
relatively flat spectrum, with $\alpha = -0.3$. 
The mean spectral index for normal pulsars 
is $-1.5$ and fewer than 10\% have $\alpha > -0.5$ 
(Lorimer et al.~1995). 
Of the two currently 
known radio-emitting magnetars, XTE1810$-$197 has $\alpha = -0.5$ 
in the frequency range $0.7 - 42$~GHz \citep{Camilo06}, and 
1E1547.0$-$5408 exhibits a flat or rising spectrum over 
$1.4 - 6.6$~GHz \citep{Camilo07}. The flat spectrum of 
PSR J1746$-$2850I in addition to its spin parameters  suggest
 that this object belongs to a class of 
young neutron stars bridging the gap between magnetars and 
canonical radio pulsars. 

\subsection{PSR J1746$-$2850II}

\begin{figure}
\begin{center}
\includegraphics[width=\linewidth]{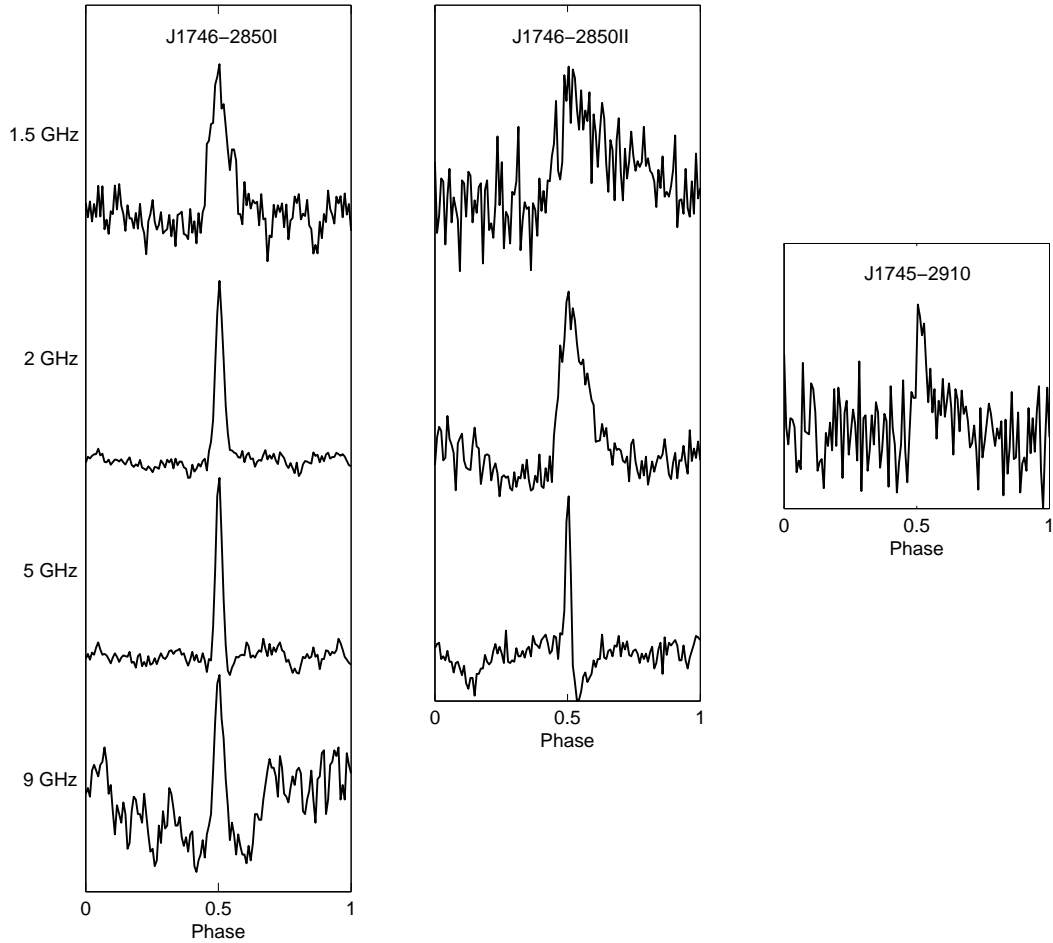}
\caption{Pulse profiles at 1.5, 2, 5, and 9~GHz for J1746-2850I (left), J1746-2850II (middle), and J1745-2910 (right). 
Scattering broadening results in an exponential tail to the pulse profiles at 2 and 1.5~GHz. J1746-2850II is scattered most severely of the three pulsars due to its high DM.\label{fig_profs}}
\end{center}
\end{figure}

J1746$-$2850II was timed simultaneously with J1746$-$2850I because the two pulsars are contained within the same beam of the GBT at 1.4 and 2 GHz. It has the highest dispersion measure of any pulsar known to date, indicating that of the five pulsars currently known within 1\degrees\ of \sgr, J1746-2850II is likely closest to \sgr\ in radial distance. It is an old pulsar, with a characteristic age $\tau_{\rm c} = 2$~Myr.  While most of the timing observations were made at 2~GHz, the receiver was unavailable in July and August 2008 so observations were made at 1.5~GHz. J1746$-$2850II is significantly scattered at this frequency (Fig.~\ref{fig_profs}). While it was detected during a 2.25~h observation in June 2008, it was too weak to detect in our routine 1-hr observations at 1.5~GHz. Consequently, its timing solution has a larger position uncertainty than that of J1746-2850I.

The pulse profile of J1746$-$2850II at 4.8~GHz is symmetric and Gaussian-like without a discernible scattering tail (Fig.~\ref{fig_profs}). We simulate the effect of scattering broadening by convolving a Gaussian matching the 4.8~GHz profile with an exponential depending on the scattering time, $\tau_{sc}$. We compare the results to the 1.5 and 2~GHz profiles using least-squares fitting and obtain $\tau_{sc} \approx 140$~ms at 2~GHz and 266~ms at 1.5~GHz. 


\subsection{PSR J1745$-$2910}


PSR J1745$-$2910 has a period of 0.982~s and DM of 1088~pc~cm${}^{-3}$. Its nominal position is $12\amin$ from \sgr\ and $\sim 8\amin$ from J1745$-$2912. Due to the large uncertainties in the positions of both pulsars, they may be between $3\amin - 14\amin$ apart. Using Eqn.~\ref{eqn_flux}, we estimate the flux density of J1745$-$2910 to be $\sim 0.2$~mJy at 2~GHz.

\subsection{PSR~J1746$-$2856 and PSR~J1745$-$2912}

The two pulsars discovered at 3.1~GHz by \cite{j+06} are in our 2~GHz search area (Fig.~\ref{fig_grid}). PSR J1746$-$2856 was blindly detected at up to $\sim 6\amin$ from its position. PSR J1745-2912 was not detected either blindly or by folding data from search pointings near its position with its period of 187~ms. The reported scattering time for this pulsar is $25 \pm 3$~ms at 3.1~GHz. If we assume a Kolmogorov scattering spectrum with a dependence of the scattering time on frequency of $\tau_{\rm sc} \propto f^{-4.0}$, the scattering time for PSR J1745-2912 at 2~GHz is 144~ms. This is close to its period and therefore we attribute the non-detection to scattering broadening. 

\section{The Galactic Center Environment}

Scattering from the dense region near \sgr\ has been modeled as
both a thin screen and as an extended scattering volume.
\cite{lc98} estimate that the scattering screen 
around the Galactic center is $133_{-80}^{+200}$~pc from \sgr. 
The NE2001 model \citep{NE2001} 
uses an ellipsoidal scattering volume with a scale height of 26~pc and characteristic radius of 145~pc that is slightly
offset from \sgr. 
The expected DM of a pulsar at the location of \sgr\ is
$\sim 1600$~pc~cm$^{-3}$ with a corresponding pulse-broadening
time  
$\sim 400$~s or $\sim 2000$~s at 1~GHz for the screen or extended models,
respectively.   Pulsars on the near side of \sgr\ will have smaller
values of DM and much smaller pulse broadening times.   

All five pulsars within 1\degree\ of \sgr\ have high DMs  that
are too large to be accounted for by 
the Galactic disk components of the  NE2001 electron density model.
If the pulsars are part of a disk population in the foreground
of \sgr\, the large DMs might
be due to an unmodeled excess of 
intervening ionized gas associated with foreground HII regions. 
However, the two \cite{j+06} pulsars are far in angular separation 
from one another and from the three pulsars we have found, making unlikely
the possibility of a single HII region contributing to their 
large DMs. Similarly unlikely is the coincidence of five out 
of five pulsars being affected by separate HII regions. 

\begin{figure}
\begin{center}
\includegraphics[width=\linewidth]{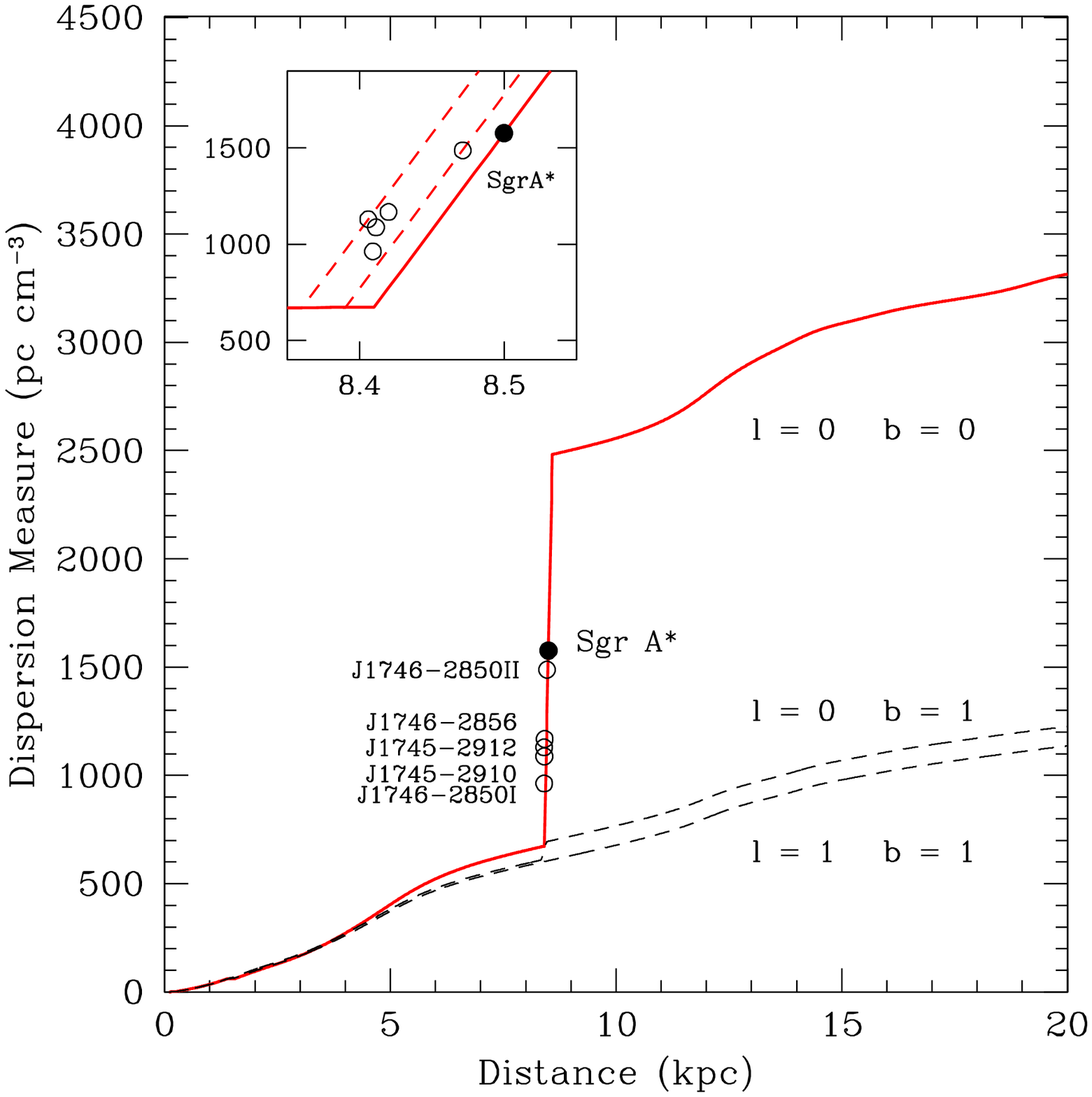}
\caption{Expected dispersion measure vs. distance for several lines of sight in the inner Galaxy based on the NE2001 model of Galactic ionized gas \citep{NE2001}. The inset shows expected DM vs. distance for the lines of sight to the five known pulsars within 1\degrees\ of \sgr. The expected DM for objects physically near \sgr\ is $\gtrsim 1500$~pc~cm$^{-3}$. The vertical part of the curve for a line of sight towards the Galactic center corresponds to a dense scattering screen around \sgr\ which makes detection of pulsars in this region difficult \citep{lc98}.\label{fig_dmvsd}}
\end{center}
\end{figure}

In the absence of any unmodeled HII regions, the 
NE2001 model places 
all five pulsars within $\sim 100$~pc of \sgr\ 
using the DM values (Fig.~\ref{fig_dmvsd}). 
However the predicted pulse broadening times for the
pulsars are several orders of magnitude larger than the measured
value for J1746$-2850$II and the upper bounds on the other objects.  
This signifies that the near-side boundary and/or density of the idealized 
Galactic-center component of the model are incorrect or
that the scattering volume is 
patchy, as suggested by  
\cite{Lazio99}.   In either case, like any distance calculated using
DM, the distances of the five pulsars from \sgr\  
are highly model dependent and cannot be refined to better than 
$\sim \pm 100$~pc. However, unless there is an 
electron density component in the foreground disk (or spiral arm)
of the Galaxy, the large DMs imply that the pulsars are no more than
about  200~pc from \sgr.  Galactic center pulsars will be key for
improving the Galactic center component of the next electron density
model, NE2008 (Cordes et al., in preparation).   



\subsection{The Arches and Quintuplet Clusters}

Could J1746$-$2850I have been born in either the Arches or Quintuplet cluster?  Both clusters contain numerous massive stars, which are neutron star progenitors. \cite{kfkn06} find the present day mass function of the Arches cluster to extend to at least 40~\msun, while \cite{fng+02} find the Arches cluster to contain stars exceeding 100~\msun\ with the age of the cluster estimated to be $2\pm1$~Myr, old enough for neutron stars to have formed. Current estimates for the maximum progenitor mass that will produce a neutron star are 20~\msun\ for solitary progenitors and as high as $50 - 80$~\msun\ for binary progenitors due to mass loss in Roche lobe overflow \citep{bt08}. The Quintuplet cluster has an estimated age of $\sim 4$~Myr \citep{Figer99} but contains the Pistol Star, whose mass of 150~\msun\ points to an age no larger than 2~Myr. These estimates indicate that the Quintuplet cluster is also old enough to have already formed neutron stars. 


The current position estimate for J1746$-$2850I places it within 2\amin\ of the Quintuplet cluster (Fig.~\ref{fig_grid}). If the cluster and the pulsar are roughly 8.5~kpc from Earth, this angle corresponds to a distance of 5~pc. A transverse velocity of 375~km/s would have allowed the pulsar to reach its present location from the Quintuplet cluster within 13~kyr. 
Our double discovery in a single pointing suggests that the Arches and Quintuplet region may have a relative excess of pulsars. 


\subsection{Galactic Center Pulsar Population}

We show here that the pulsar detections within 15 arcmin of \sgr\ imply the existence of a Galactic center subpopulation of pulsars and that this population may be quite sizable. First, given the small solid-angle coverage of our survey, $\Delta\Omega \approx 10^{-4.2}$~sr, we do not expect to find {\it any} pulsars from the Galactic disk population. Integrating either a simple disk model with a uniform density or using the spatial distribution reported by \citet{Lorimer06}, we find that only 1 to 2 disk pulsars beamed toward us are expected in the volume between Earth and \sgr.  When we include the survey sensitivity, which depends on the pulsar period, and the luminosity and period distributions, the expected number of detected pulsars $\ll 1$.

\begin{figure}
\includegraphics[width=\linewidth]{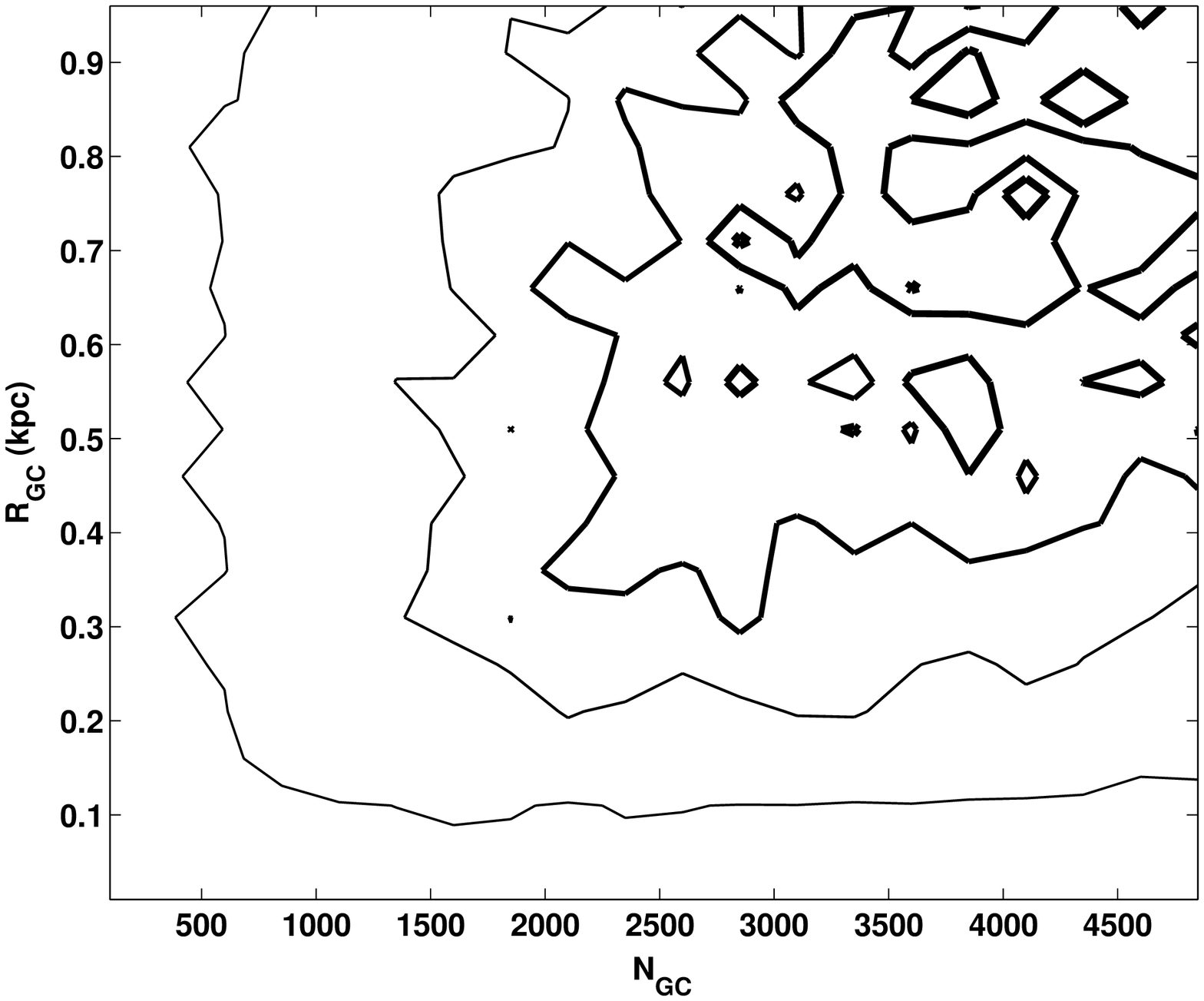}
\caption{Likelihood contours based on the Poisson probability of detecting on average $\left<N\right>$ pulsars of the simulated population in a survey beam where there are $k$ actual detections. Contours from left to right and in order of increasing thickness correspond to ${\cal L} = (0.1, 1, 2, 4, 6, 8) \times 10^{-6}$.}\label{fig_lkl}
\end{figure}

To assess the rough parameters of a Galactic center pulsar population, we employed a simple modeling approach. We used an ellipsoidal distribution with characteristic radius in the plane of the Galaxy $R_{\rm GC}$ that contains $N_{\rm GC}$ pulsars. The scale height of the population was fixed at $H_{\rm GC} = 0.026$~kpc, equal to that of the scattering region in the NE2001 model. Through Monte Carlo, we generated pulsars consistent with this spatial distribution and drawn from populations with the period and luminosity distributions of \citet{Lorimer06}. Detection criteria were applied using the GBT observation parameters reported earlier and the Parkes survey parameters reported by \citet{j+06}, and taking into account the effects of pulse broadening from both residual dispersion smearing and scattering on the harmonic sum. As a test statistic we used the likelihood function ${\cal L} =\prod_i P(l_i,b_i,\left<N_i\right>,k_i)$. A term in the product corresponds to the Poisson probability of detecting on average $\left<N\right>$ pulsars of the simulated population in the $i$-th survey beam, where there are $k$ actual detections, $P(\left<N\right>,k) = \left<N\right>^k e^{-\left<N\right>}/k!$. The population was generated 1000 times for each $N_{\rm GC},R_{\rm GC}$ pair. Figure~\ref{fig_lkl} shows contours of ${\cal L}$, which indicate that $N_{\rm GC} \gtrsim 2000$ and $R_{\rm GC} \gtrsim 0.3$~kpc are lower bounds on parameter values. 


Extending our grid up to $N_{GC} = 10^4$ and $R_{GC} = 5$~kpc, we found no upper bounds on the parameters. We expect such bounds from a more comprehensive analysis that includes other survey results over a larger region, which we defer to another paper. 


\section{Conclusions}

We have discovered three pulsars within 12\amin\ of \sgr. The low ($\ll 1$) number of detectable disk pulsars in our survey volume and the high dispersion measures of the new pulsars indicate that they are within the dense scattering region surrounding \sgr\ and part of a neutron star population associated with the Galactic center. Based on a Monte Carlo simulation of this population which incorporates our survey results as well as those of \cite{j+06}, we conclude that there are $N_{GC} \gtrsim 2000$ active pulsars associated with the Galactic center. Based on the average lifetime of canonical pulsars, $\sim 10$~Myr, we obtain a rough estimate of the radio pulsar birth rate in the Galactic center of $\gtrsim 2 \times 10^{-4}$~yr$^{-1}$. Considering that this is a lower limit, and that not all supernovae produce radio pulsars, this result is consistent with the rate of Galactic center supernovae ($\sim 10^{-3}$~yr$^{-1}$) based on a survey of compact radio sources \citep{lc08} and observations of soft X-ray emitting plasma \citep{muno04}.

\section{Acknowledgments}
We thank Paulo Freire and Scott Ransom for helpful discussions. The GBT is a telescope operated by the National Radio Astronomy Observatory (NRAO), a facility of the National Science Foundation operated under cooperative agreement by Associated Universities, Inc. J.S.D. was partially funded by an NRAO Student Support grant during this work. Basic research in radio astronomy at the NRL is supported by 6.1 Base funding. This work is supported at Cornell by NSF grant AST0807151.

\end{document}